\documentclass[prl,twocolumn,superscriptaddress]{revtex4}
\usepackage{amssymb,amsmath,amsthm,graphicx}
\usepackage{latexsym}
\usepackage{bm}
\usepackage[]{hyperref} % to have online links

\pdfoutput=1

\def\be{\begin{equation}}
\def\ee{\end{equation}}
\def\beq{\begin{eqnarray}}
\def\eeq{\end{eqnarray}}

\newcommand{\lp}{\left(}
\newcommand{\rp}{\right)}

\begin{document}

\def\lsim{\mathrel{\rlap{\lower4pt\hbox{\hskip1pt$\sim$}}
    \raise1pt\hbox{$<$}}}
\def\gsim{\mathrel{\rlap{\lower4pt\hbox{\hskip1pt$\sim$}}
    \raise1pt\hbox{$>$}}}
%\def\sqr#1#2{{\vcenter{\vbox{\hrule height.#2pt
%         \hbox{\vrule width.#2pt height#1pt \kern#1pt
%         \vrule width.#2pt}
%         \hrule height.#2pt}}}}
%\def\square{\mathchoice\sqr66\sqr66\sqr{2.1}3\sqr{1.5}3}
%%%%%%%%%%%%%%%%%%%%%%%%%%%%%%
\def\be{\begin{equation}}
\def\ee{\end{equation}}
\def\bea{\begin{eqnarray}}
\def\eea{\end{eqnarray}}
%%%%%%%%%%%%%%%%%%%%%%%%%%%%%%
\newcommand{\der}[2]{\frac{\partial{#1}}{\partial{#2}}}
\newcommand{\dder}[2]{\partial{}^2 #1 \over {\partial{#2}}^2}
\newcommand{\dderf}[3]{\partial{}^2 #1 \over {\partial{#2} \partial{#3}}}
\newcommand{\eq}[1]{Eq.~(\ref{eq:#1})}
\newcommand{\dd}{\mathrm{d}}

\title{AdS nonlinear instability: moving beyond spherical symmetry}

\author{ \'Oscar J. C. Dias}
\email{ojcd1r13@soton.ac.uk}
\affiliation{STAG research centre and Mathematical Sciences, University of Southampton, UK} 
\author{Jorge E. Santos}
\email{jss55@cam.ac.uk}
\affiliation{Department of Applied Mathematics and Theoretical Physics, University of Cambridge, Wilberforce Road, Cambridge CB3 0WA, UK}

\begin{abstract}
Anti-de Sitter (AdS) is conjectured to be nonlinear unstable to a weakly turbulent mechanism that develops a cascade towards high frequencies, leading to black hole formation \cite{DafermosHolzegel2006,Bizon:2011gg}. We give evidence that the gravitational sector of perturbations behaves differently from the scalar one studied in \cite{Bizon:2011gg}. In contrast with \cite{Bizon:2011gg}, we find that not all gravitational normal modes of AdS can be nonlinearly extended into periodic horizonless smooth solutions of the Einstein equation. In particular, we show that even seeds with a single normal mode can develop secular resonances, unlike the spherically symmetric scalar field collapse studied in \cite{Bizon:2011gg}. Moreover, if the seed has two normal modes, more than one resonance can be generated at third order, unlike the spherical collapse of \cite{Bizon:2011gg}. We also show that  weak turbulent perturbative theory predicts the existence of direct and inverse cascades, with the former dominating the latter for equal energy two-mode seeds.   
\end{abstract}

\maketitle

%%%%%%%%%%%%%%%%%%%%%%%%%%%%%%%%%%%%%%%%%%%%%%%%%%%%%%%%%%%%%%%%%%%%%%%%%%%
%%%%%%%%%%%%%%%%%%%%%%%%%%%%%%%%%%%%%%%%%%%%%%%%%%%%%%%%%%%%%%%%%%%%%%%%%%%

%%%%%%%%%%%%%%%%%%%%%%%%%%%%%%%%%%%%%%%%%%%%%%%%%%%%%%%%%%%%%%%%%%%%%%%%%%%
%\section{Introduction}
%%%%%%%%%%%%%%%%%%%%%%%%%%%%%%%%%%%%%%%%%%%%%%%%%%%%%%%%%%%%%%%%%%%%%%%%%%%
{\bf 1.~Introduction ---} For many years, Anti-de Sitter (AdS) spacetime was regarded as the odd cousin of de-Sitter and Minkowski spacetimes. However, with the dawn of AdS/CFT \cite{Maldacena:1997re}, studying gravitational dynamics in AdS become more than a mere academic exercise.

A topic that has attracted particular attention over the last years is the issue of global nonlinear stability of AdS \cite{DafermosHolzegel2006,Bizon:2011gg,Dias:2011ss,Dias:2012tq,Buchel:2012uh,Buchel:2013uba,Maliborski:2013jca,Bizon:2013xha,Maliborski:2012gx,Maliborski:2013ula,Baier:2013gsa,Jalmuzna:2013rwa,Basu:2012gg,2012arXiv1212.1907G,Friedrich:2014raa,Maliborski:2014rma,Abajo-Arrastia:2014fma,Balasubramanian:2014cja,Bizon:2014bya,Balasubramanian:2015uua,daSilva:2014zva,Craps:2014vaa,Basu:2014sia,Okawa:2014nea,Deppe:2014oua,Dimitrakopoulos:2014ada,Buchel:2014xwa,Craps:2014jwa,Basu:2015efa,Yang:2015jha,Okawa:2015xma,Bizon:2015pfa,Dimitrakopoulos:2015pwa,Green:2015dsa,Deppe:2015qsa,Craps:2015iia,Craps:2015xya,Evnin:2015gma,Menon:2015oda,Jalmuzna:2015hoa,Evnin:2015wyi,Freivogel:2015wib}, which was ignited by the seminal work of Bizon and Rostworowski in \cite{Bizon:2011gg}. This instability was first conjectured by Dafermos and Holzegel in \cite{DafermosHolzegel2006}. While it would be desirable to study the nonlinear stability of AdS with no symmetry restrictions, this problem seems intractable both from an analytical and numerical standpoints. To circumvent this, the authors of \cite{Bizon:2011gg} restricted themselves to spherical symmetry. However, since there are no gravitational spherical waves within pure Einstein's theory of gravity, in \cite{Bizon:2011gg} a massless scalar field was added, which essentially controls all the dynamics of the system.

The numerical results of \cite{Bizon:2011gg} suggest that AdS is nonlinearly unstable to the formation of an arbitrarily small black hole, whose mass is controlled by the energy contained in the initial data - a weakly turbulent instability. This is to contrast with Minkowski and de Sitter spacetimes, for which the nonlinear stability problem has long been proved \cite{Christodoulou:1993uv,friedrich86}. While this nonlinear instability seems to occur for generic perturbations, there are perturbations which do not necessarily lead to an instability \cite{Dias:2012tq,Buchel:2012uh,Maliborski:2013jca,Dimitrakopoulos:2015pwa,Green:2015dsa,Craps:2015xya}, leading to the existence of islands of stability. How large are these islands as a function of the amplitude of the initial data, is a question that has not yet been fully understood. There is however, a one-parameter family of initial data that seems to be particularly important: time-periodic solutions of the Einstein equation with a negative cosmological constant. They correspond to nonlinear extensions of normal modes of spherically symmetry scalar fields in AdS, and were coined `oscillons' \footnote{Similar solutions exists in the context of complex scalar fields, and are called `boson stars'.}.

In \cite{Bizon:2011gg}, not only the problem was analysed from a numerical standpoint, but also an analytic method was proposed to detect the onset of such an instability. The idea is to use standard perturbation theory to third order in the amplitude of the linear seed, which we schematically denote by $\varepsilon$. At third oder in $\varepsilon$, a secular term of the form $t\,\varepsilon^3$ was found, which invalidates standard perturbation theory for timescales $t\geq \varepsilon^{-2}$. This secular growth occurs because the linear spectrum is fully commensurable and, as such, nonlinearities can create resonances. This analytic approach was shortly afterwards generalised for certain gravitational perturbations \cite{Dias:2011ss}, even though a more systematic approach for the gravitational perturbations has never been done.

There are modifications of standard perturbation theory that can capture up to time scales $t\lesssim \varepsilon^{-2}$. Such schemes can be shown to be all equivalent, and go under the following names: two time scale formalism \cite{Balasubramanian:2014cja}, renormalisation group perturbation methods \cite{Craps:2014vaa,Craps:2014jwa} and resonant approximation \cite{Bizon:2015pfa}. They all rely on the assumption that resonant modes appear in a single sector of perturbations and that only a specific type of resonant modes is relevant. We will see later that it is not clear whether this is the case for gravitational perturbations.

While we do not know if any solution of the Einstein equation with a fully resonant spectrum necessarily possesses a nonlinear instability, it is clear it is a necessary condition for the existence of the weakly turbulent instability, as shown in \cite{Dias:2012tq}. In this letter we perform third order perturbation theory calculations for a plethora of seeds, and find that the gravitational case is more rich than the spherically symmetry cases analysed so far. Our results indicate that the spherically symmetric scalar field collapse is unlikely to be a good toy model for the general gravitational collapse.
\\
%%%%%%%%%%%%%%%%%%%%%%%%%%%%%%%%%%%%%%%%%%%%%%%%%%
%\section{2.~Perturbation theory:}
%%%%%%%%%%%%%%%%%%%%%%%%%%%%%%%%%%%%%%%%%%%%%%%%%%
{\bf 2.~Perturbation theory ---}
Consider Einstein-AdS$_4$ theory with action and equation of motion, 
\begin{equation}
S=\int d^4x\;\sqrt{-g}\left(R+\frac{6}{L^2}\right), \quad R_{\mu\nu}+\frac{3}{L^2}g_{\mu\nu}=0\,,
\label{eq:action}
\end{equation}
where $L$ is the AdS length scale. Within perturbation theory with expansion parameter $\varepsilon$, we perturb this equation of motion about the global AdS background $\bar{g}$  as  $g = \bar{g}+\sum_k h^{(k)}\varepsilon^k$ (more details can be found in the Supplementary Material, in \cite{Dias:2011ss} and in the companion manuscript \footnote{In a companion paper we  give an exhaustive description of the technical machinery required to get our results. We also present the quantum numbers of all the harmonic modes (in total hundreds of them) that are excited in the examples described here.}). At each order  in perturbation theory, the Einstein equations yield
\begin{equation}
\Delta_L h_{ab}^{(k)} = T^{(k)}_{ab},
\label{eq:perturb}
\end{equation}
where $\Delta_L$ is a second order operator constructed uniquely from $\bar{g}$ and $T^{(k)}$ is a function of $\{h^{(j\leq i-1)}\}$ and their derivatives. 

We are interested in finding regular finite energy and angular momentum solutions of (\ref{eq:perturb}) for $k\geq 1$. We explore the $SO(3)$ symmetry of AdS and use the  Kodama-Ishibashi formalism to solve \eqref{eq:perturb} \cite{Kodama:2003jz,Kodama:2003kk}. Namely, we expand the source $T^{(k)}_{ab}$ (and thus $h_{ab}^{(k)}$) as a sum of scalar $\mathcal{T}^{({\rm \bf s})}_{\ell_{\rm \bf s},m_{\rm \bf s}}$ and vector $\mathcal{T}^{({\rm \bf v})}_{\ell_{\rm \bf v},m_{\rm \bf v}}$ type modes. These are characterised by the usual spherical harmonic quantum numbers $\ell$ and $m$. It follows that any solution of (\ref{eq:perturb}) is described by two decoupled PDEs of the form \cite{Kodama:2003jz,Kodama:2003kk}
\begin{equation}
\Box_2 \Phi^{(k)}_{\ell,m}(t,r)+V^{(k)}_{\ell}(r) \Phi^{(k)}_{\ell,m}(t,r)=\widetilde{\mathcal{T}}^{(k)}_{\ell,m}(t,r),
\label{eq:master}
\end{equation}
where $V^{(k)}_\ell(r)$ is a potential that only depends on $\ell$ and $\Box_2$ is the d'Alambertian of the two dimensional orbit space spanned by $(t,r)$. One of the equations governs scalar-type modes and the other vector-type modes. $\Phi^{(k)}_{\ell,m}(t,r)$ is a gauge invariant variable from which  $h^{(k)}_{\ell,m}$ $-$ that solves (\ref{eq:perturb}) $-$ can be uniquely recovered (in a particular gauge)  through a linear differential map \cite{Kodama:2003jz}. $\widetilde{\mathcal{T}}^{(k)}_{\ell,m}(t,r)$ is a scalar source term, that can be expressed as a function of the components of $\mathcal{T}^{(k)}_{\ell,m}$ and its derivatives.

We require the solution to be asymptotically global AdS and everywhere regular (including the origin). The time dependence of our source term is an overall multiplicative factor of the form $\cos(\omega  t)$, so (\ref{eq:master}) is solved assuming the separation {\it ansatz} $\Phi^{(k)}_{\ell,m}(t,r)=\cos(\omega  t) R(r)$. There is however an exception, namely when the source frequency $\omega$ coincides with one of the AdS gravitational normal mode frequencies  $\bar{\omega}$  (\emph{i.e.} the characteristic oscillation frequencies of the homogeneous AdS background to be described below). If this is the case we say that the modes are \emph{resonant} and the general solution of (\ref{eq:master}) now includes a solution that grows linearly with time:
\begin{equation}
\Phi^{(k)}_{\ell,m}(t,r)=\cos(\omega t)R_{\omega,\ell,m}(r)+t\,\sin(\omega t)L_{\omega,\ell,m}(r).
\label{eq:resonant}
\end{equation}

%%%%%%%%%%%%%%%%%%%%%%%%%%%%%%%%%%%%%%%%%%%%%%%%%%
%\section{3. Geons:}
%%%%%%%%%%%%%%%%%%%%%%%%%%%%%%%%%%%%%%%%%%%%%%%%%%
{\bf 3.~Normal modes without a nonlinear extension and geons ---}
At first order, $\widetilde{\mathcal{T}}^{(1)}_{\ell,m}\equiv0$, and the AdS background is time-independent so one can Fourier transform \eqref{eq:master} in time, with Fourier parameter $\bar{\omega}_{\ell,p}$, reducing the problem to the study of a single ODE of the St\"urm-Liouville type.
Consider first the scalar-type modes. Requiring regularity at the origin and the solution to be asymptotically globally AdS quantises the normal mode frequencies as 
 \begin{equation}  \label{spectrumS}
\bar{\omega}_{\ell_{\rm \bf s},p_{\rm \bf s}}L= 1+\ell_{\rm \bf s} +2\,p_{\rm \bf s}, \quad \hbox{where}\quad p_{\rm \bf s}\in\{0,1,2\ldots \}
  \end{equation}
is the radial overtone. The associated regular orthogonal basis of eigenfunctions for $\Phi^{(1)}_{\ell,m}(t,r)$ that solves (\ref{eq:master}) is 
\begin{eqnarray}
\label{basisS}
&& \hspace{-0.5cm} \left\{ e_{\ell_{\rm \bf s},p_{\rm \bf s}} \right\}= {\biggl \{}  \tilde{a}_{\ell_{\rm \bf s},p_{\rm \bf s}} \,\frac{r^{\ell_{\rm s}+1}} {\left(r^2+L^2\right)^{\frac{\ell_{\rm s}+1 }{2}}}  \times \\
&&\hspace{-0.2cm}
_2F_1\!\!\left(\frac{\ell_{\rm s} -\bar{\omega}_{\ell_{\rm \bf s},p_{\rm \bf s}} L +1}{2},\frac{\ell_{\rm s} +\bar{\omega}_{\ell_{\rm \bf s},p_{\rm \bf s}} L +1}{2};\frac{1}{2};\frac{L^2}{r^2+L^2}\right)\!\!\!{\biggr \}},\nonumber
\end{eqnarray}
where $_2F_1$ is the Gaussian hypergeometric function and $\tilde{a}_{\ell_{\rm \bf s},p_{\rm \bf s}}$ is a normalisation real constant of our choice.

For vector-type modes the boundary conditions quantise the frequencies as 
\begin{equation}  \label{spectrumV}
\bar{\omega}_{\ell_{\rm \bf v},p_{\rm \bf v}}L= 2+\ell_{\rm \bf v} +2 \,p_{\rm \bf v}, \quad \hbox{with}\quad p_{\rm \bf v}\in\{0,1,2\ldots \}.
\end{equation}
Its associated orthogonal basis of eigenfunctions is
\begin{eqnarray}
\label{basisV}
&& \hspace{-0.5cm}\left\{ e_{\ell_{\rm \bf v},p_{\rm \bf v}} \right\}=  {\biggl \{}  \tilde{b}_{\ell_{\rm \bf v},p_{\rm \bf v}} \,\frac{r^{\ell_{\rm v}+1}} {\left(r^2+L^2\right)^{\frac{\ell_{\rm v}}{2}+1}}  \times \\
&&\hspace{-0.2cm}
 _2F_1\!\!\left(\!\frac{\ell_{\rm v} -\bar{\omega}_{\ell_{\rm \bf v},p_{\rm \bf v}} L +2}{2},\frac{\ell_{\rm v} +\bar{\omega}_{\ell_{\rm \bf v},p_{\rm \bf v}} L +2}{2};\frac{3}{2};\frac{L^2}{r^2+L^2}\!\right) \!\!\!{\biggr \}},\nonumber
\end{eqnarray}
where $\tilde{b}_{\ell_{\rm \bf v},p_{\rm \bf v}}$ is a normalisation constant. Gravitational waves are described in both sectors by $\ell \ge 2$ and $\ell\geq |m|$.
  
We  emphasize two key properties of the linearized spectrum of frequencies \eqref{spectrumS} and \eqref{spectrumV}: 1) the spectrum is purely real, and 2) the spectrum is commensurable, \emph{i.e.} the sum or difference of any two frequencies yields a frequency that still belongs to the spectrum. 

We are ready to extend our computation to higher order. In this Section we start with a linear seed consisting of a {\it single} scalar or vector mode. We will  describe many such cases and we summarise our main conclusions in Tables \ref{Table:geonsS1} and \ref{Table:geonsV}. 
In these tables, the first column describes our seed, \emph{i.e.} the quantum numbers $\{ \ell,m, p, \bar{\omega}\}$ of the particular (single) normal mode we start with. The content of the other columns is described in the caption of Table \ref{Table:geonsS1}. Table \ref{Table:geonsS1} (Table \ref{Table:geonsV}) considers the back-reaction of scalar (vector) normal modes. 

\begin{table}[ht]
\begin{eqnarray}
\nonumber
\begin{array}{||  c  ||  c  ||  c  | c  | c ||}\hline\hline
\hbox{Normal mode} &  \#    &  \#  &  \hbox{Removable} & \hbox{Secular} \\
\{\ell,m,p,\bar{\omega}\} & \hbox{modes}  &   \hbox{modes}   &\hbox{resonance} & \hbox{resonances } \\
\hbox{ at } \mathcal{O} \lp \varepsilon \rp  &  \mathcal{O} \lp \varepsilon^2 \rp & \mathcal{O} \lp \varepsilon^3 \rp  & \hbox{$\lp \: -L\,\omega^{(2)} \: \rp$} & \{\ell,m,p,\omega\} \\
\hline \hline
{\bf \{2,0,0,\frac{3}{L}\}_{\rm \bf s}} & 6_{\rm \bf s} & 8_{\rm \bf s} &  \{2,0,0,\frac{3}{L}\}_{\rm \bf s} & \hbox{None} \\
     &  0_{\rm \bf v} & 0_{\rm \bf v} &  \lp \: \frac{3663}{8960} \: \rp & \hbox{\bf (Geon ?)}  \\
 \hline
\{2,0,1,\frac{5}{L}\}_{\rm \bf s}   & 6_{\rm \bf s} & 8_{\rm \bf s} &  \{2,0,1,\frac{5}{L}\}_{\rm \bf s} & \{4,0,0,\frac{5}{L}\}_{\rm \bf s} \\
                              & 0_{\rm \bf v} & 0_{\rm \bf v} &  \lp \: \frac{34397}{5376}\:  \rp &   \\
\hline
 \{4,0,0,\frac{5}{L}\}_{\rm \bf s} & 10_{\rm \bf s} & 14_{\rm \bf s} &  \{4,0,0,\frac{5}{L}\}_{\rm \bf s} & \{2,0,1,\frac{5}{L}\}_{\rm \bf s}  \\
     &  0_{\rm \bf v} & 0_{\rm \bf v} &  \lp \: \frac{52311625}{21446656} \: \rp &   \\
 \hline
{\bf \{2,1,0,\frac{3}{L}\}_{\rm \bf s} }   & {5}_{\rm \bf s} & {5}_{\rm \bf s} & \{2,1,0,\frac{3}{L}\}_{\rm \bf s} & \hbox{None} \\ 
    & {2}_{\rm \bf v} & {4}_{\rm \bf v} & \lp \:  \frac{123}{64}  \:  \rp  &  \hbox{\bf (Geon ?)}  \\ 
\hline
{\bf \{2,2,0,\frac{3}{L}\}_{\rm \bf s} }  & 4_{\rm \bf s} & 4_{\rm \bf s} & \{2,2,0,\frac{3}{L}\}_{\rm \bf s} & \hbox{None} \\
                               & 2_{\rm \bf v} & 2_{\rm \bf v} & \lp\: \frac{14703}{1120} \:\rp  & \hbox{\bf (Geon)}  \\
 \hline
\{2,2,1,\frac{5}{L}\}_{\rm \bf s}  & 4_{\rm \bf s} & 4_{\rm \bf s} &  \{2,2,1,\frac{5}{L}\}_{\rm \bf s} &  \{4,2,0,\frac{5}{L}\}_{\rm \bf s} \\
               & 2_{\rm \bf v} & 2_{\rm \bf v} & \lp\:  \frac{9409723}{70560} \: \rp & \{3,2,\frac{5}{L}\}_{\rm \bf v} \\
 \hline
{\bf \{3,3,0,\frac{4}{L}\}_{\rm \bf s} }   & {5}_{\rm \bf s} & {5}_{\rm \bf s} & \{3,3,0,\frac{4}{L}\}_{\rm \bf s} & \hbox{None} \\ 
    & {3}_{\rm \bf v} & {3}_{\rm \bf v} & \lp \:  \frac{27881625}{32032}  \:  \rp  &  \hbox{\bf (Geon)}  \\ 
\hline 
\{3,2,0,\frac{4}{L}\}_{\rm \bf s}   & 5_{\rm \bf s} & 6_{\rm \bf s} &  \{3,2,0,\frac{4}{L}\}_{\rm \bf s}  &  \{2,2,0,\frac{4}{L}\}_{\rm \bf v}  \\
                               & 2_{\rm \bf v} & 5_{\rm \bf v} & \lp\:  \frac{8081875}{72072} \:\rp &   \\
 \hline
{\bf \{4,4,0,\frac{5}{L}\}_{\rm \bf s} }   &  6_{\rm \bf s} & 6_{\rm \bf s} &  \{4,4,0,\frac{5}{L}\}_{\rm \bf s} & \hbox{None} \\
                               & 4_{\rm \bf v} & 4_{\rm \bf v} &  \lp\: \frac{7010569125}{77792} \:\rp & \hbox{\bf (Geon)} \\
 \hline
\{4,2,0,\frac{5}{L}\}_{\rm \bf s}   &   8_{\rm \bf s} & 10_{\rm \bf s}   &  \{4,2,0,\frac{5}{L}\}_{\rm \bf s}  & \{2,2,1,\frac{5}{L}\}_{\rm \bf s} \\
&  4_{\rm \bf v} & 7_{\rm \bf v} & \lp \frac{163492329375}{243955712} \rp   &  \{3,2,0,\frac{5}{L}\}_{\rm \bf v} \\
  \hline
{\bf \{6,6,0,\frac{7}{L}\}_{\rm \bf s} }  &   8_{\rm \bf s} & 8_{\rm \bf s}  &   \{6,6,0,\frac{7}{L}\}_{\rm \bf s} & \hbox{None} \\
          &  6_{\rm \bf v} & 6_{\rm \bf v}  &   \lp -L\,\omega^{(2)}_{6,6,0} \rp  & \hbox{\bf (Geon)}  \\
\hline\hline
\end{array}
\end{eqnarray}
\caption{Back-reaction of seeds with a single {\it scalar} normal mode. The first column describes the quantum numbers $\{ \ell,m, p, \bar{\omega}\}_{\rm \bf j}$ (${\rm \bf j} \in \{\rm \bf s,v \}$) of the normal mode we start with at linear order. Each row describes a distinct case. The second and third columns display the number ($ \# $) of scalar (${\rm s}$) and vector (${\rm v}$) harmonics that are excited at $2^{\rm nd}$ and $3^{\rm rd}$ order, respectively. In the fourth column we identify the quantum numbers $\{ \ell,m, p, \omega\}_{\rm \bf j}$ of the third order removable resonance. It is also given the frequency correction $-L\,\omega^{(2)}$ that removes it. The last column identifies the secular resonances  $\{ \ell,m, p, \omega\}_{\rm \bf j}$.  (In the last row, $-L\,\omega^{(2)}_{6,6,0}= \frac{8231910851500875}{3090464}$).} \label{Table:geonsS1}
\end{table}

\begin{table}[ht]
\begin{eqnarray}
\nonumber
\begin{array}{||  c  ||  c  ||  c  | c  | c ||}\hline\hline
\hbox{Normal mode} &  \#    &  \# &  \hbox{Removable} & \hbox{Secular} \\
\{\ell,m,p,\bar{\omega}\} & \hbox{modes}  &   \hbox{modes}   &\hbox{resonance} & \hbox{resonances } \\
\hbox{ at } \mathcal{O} \lp \varepsilon \rp  &  \mathcal{O} \lp \varepsilon^2 \rp & \mathcal{O} \lp \varepsilon^3 \rp  & \hbox{$\lp \: -L\,\omega^{(2)} \: \rp$} & \{\ell,m,p,\omega\} \\
\hline \hline
{\bf \{2,0,0,\frac{4}{L}\}_{\rm \bf v} } & 6_{\rm \bf s} & 0_{\rm \bf s} &   \{2,0,0,\frac{4}{L}\}_{\rm \bf v} & \hbox{None} \\
                    &  0_{\rm \bf v} & 6_{\rm \bf v} &  \lp\: \frac{1469}{26880}  \:\rp &  \hbox{\bf (Geon ?)} \\
 \hline
\{2,0,1,\frac{6}{L}\}_{\rm \bf v}  &6_{\rm \bf s} & 0_{\rm \bf s}  &   \{2,0,0,\frac{6}{L}\}_{\rm \bf v} &  \{4,0,0,\frac{6}{L}\}_{\rm \bf v} \\
                    &   0_{\rm \bf v} & 6_{\rm \bf v}  &  \lp\: \frac{19081}{376320} \:\rp &   \\
 \hline
\{2,1,0,\frac{4}{L}\}_{\rm \bf v}   & {5}_{\rm \bf s} & {4}_{\rm \bf s} & \{2,1,0,\frac{4}{L}\}_{\rm \bf v} &  \{3,1,0,\frac{4}{L}\}_{\rm \bf s}  \\ 
    & {2}_{\rm \bf v} & {5}_{\rm \bf v} & \lp \:  \frac{72361}{322560}  \:  \rp  &    \\ 
\hline 
\{2,2,0,\frac{4}{L}\}_{\rm \bf v}  & 4_{\rm \bf s} & 2_{\rm \bf s} &  \{2,2,0,\frac{4}{L}\}_{\rm \bf v}  &  \{3,2,0,\frac{4}{L}\}_{\rm \bf s} \\
               &   2_{\rm \bf v} & 4_{\rm \bf v}   &  \lp\: \frac{1247}{1008} \:\rp   &   \\
 \hline
\{3,2,0,\frac{5}{L}\}_{\rm \bf v}  & 5_{\rm \bf s} & 5_{\rm \bf s} & \{3,2,0,\frac{5}{L}\}_{\rm \bf v} & \{2,2,1,\frac{5}{L}\}_{\rm \bf s} \\
           &  3_{\rm \bf v} & 6_{\rm \bf v}  & \lp\: \frac{31995875}{4612608} \:\rp  &  \{4,2,0,\frac{5}{L}\}_{\rm \bf s}   \\
 \hline
\{7,6,0,\frac{9}{L}\}_{\rm \bf v}  &  10_{\rm \bf s} & 9_{\rm \bf s} &  \{7,6,0,\frac{9}{L}\}_{\rm \bf v}   & \{6,6,1,\frac{9}{L}\}_{\rm \bf s} \\
&   7_{\rm \bf v} & 8_{\rm \bf v} &  \lp -L\,\omega^{(2)}_{7,6,0} \rp  & \{8,6,0,\frac{9}{L}\}_{\rm \bf s} \\
\hline\hline
\end{array}
\end{eqnarray}
\caption{Back-reaction of a seed with a single {\it vector} normal mode. The information is displayed as in Table \ref{Table:geonsS1}. The last frequency correction is  $-L\,\omega^{(2)}_{7,6,0}= \frac{8548214990390361}{19124224}$.} \label{Table:geonsV}
\end{table}

For all the cases, at second order $k=2$ the solution is asymptotically global AdS and regular everywhere without introducing secular terms.

Consider now the third order, $k=3$. Some of the excited harmonics are resonant: they have quantum numbers $\{ \ell,m,p,\omega \}$ such that their frequency $\omega$ matches one of the normal mode frequencies $\bar{\omega}$ in \eqref{spectrumS} or \eqref{spectrumV}. They are all of the type $\omega=2\bar{\omega}-\bar{\omega}$. Out of these, and for all the seed cases, there is a harmonic whose quantum numbers $\{ \ell,m,p,\omega \}$ coincide with those ($\{\ell,m,p,\bar{\omega}\}$) we started with in the seed. Such case is a {\it removable resonance} since we can introduce a Poincar\'e-Lindstedt frequency correction $\omega^{(2)}$ at order $\mathcal{O}(\varepsilon^2)$ (see e.g. \cite{Hunter:2004}),  
\begin{equation}\label{O2:wexpansion}
 \omega = \bar{\omega}+ \varepsilon^2 \omega^{(2)}+\mathcal{O}(\varepsilon^4),
\end{equation}
that we can then choose to eliminate the secular contribution in \eqref{eq:resonant}, while keeping the solution regular at the origin and asymptotically global AdS. These frequency corrections are  in the fourth column of Tables  \ref{Table:geonsS1} and \ref{Table:geonsV}.

The situation is different for those resonances ({\it if} present) whose quantum numbers do not match the linear seed, which are listed in the last column of Tables  \ref{Table:geonsS1} and \ref{Table:geonsV}. In this case we have secular growth because the solution cannot be made regular at the origin without introducing the linearly growing amplitude contribution in \eqref{eq:resonant} \footnote{We are not aware of a resummation procedure that removes these resonances without promoting the mode amplitudes to be functions of time, which is not periodic. That is to say, the normal mode solution cannot be back-reacted to yield a time-periodic regular nonlinear soliton.}.

There are however a `few' (although a countable infinite number of) exceptions to this scenario. Indeed, a few normal modes can be back-reacted to $\mathcal{O}(3)$ and rendered regular without introducing secular resonances. These are the boldface modes in Tables  \ref{Table:geonsS1} and \ref{Table:geonsV} \footnote{We computed their energy $E$ and angular momentum $J$ to fourth order in $\varepsilon$ and checked they obey the first law of thermodynamics, $\mathrm{d}E=\frac{\omega}{m}\mathrm{d}J +\mathcal{O}(\varepsilon^4)$. The physical interpretation for the expansion parameter $\varepsilon$ also follows from this computation: $\varepsilon$ is proportional to the square root of the energy.}. In the $\{ \ell,m,p,\bar{\omega} \}_{\rm s}=\{2,2,0,3/L\}$ case we have explicitly checked that there are no secular resonances also at  $\mathcal{O}(5)$ \cite{Dias:2011ss} and actually at any order (if we introduce frequency corrections at each even order, $\omega= \bar{\omega}+\sum_{j=1} \varepsilon^{2j}  \omega^{(2j)}$) since the full nonlinear solution has constructed numerically in \cite{Horowitz:2014hja}. The structure of the problem indicates this should hold for the other cases $\ell_{\rm s}=m_{\rm s}\geq 2$ and $p_{\rm s}=0$.
Such a gravitational normal mode that can be back-reacted to yield a nonlinear horizonless regular solution is called a {\it geon}.  It is invariant under a Killing vector which is $K\equiv\partial_t+\frac{\omega}{m}\partial_\phi$. Thus, it is not time symmetric neither axisymmetric but time-periodic \footnote{The exception are the two cases in Tables \ref{Table:geonsS1} and \ref{Table:geonsV} with $m=0$ which are axisymmetric but time-dependent.}.  

%%%%%%%%%%%%%%%%%%%%%%%%%%%%%%%%%%%%%%%%%%%%%%%%%%
%\section{4.~Direct and inverse turbulent cascades:}
%%%%%%%%%%%%%%%%%%%%%%%%%%%%%%%%%%%%%%%%%%%%%%%%%%
{\bf 4.~Direct and inverse turbulent cascades ---} In this section we show that the weakly perturbative turbulent mechanism predicts the existence of direct but also inverse frequency cascades to be observed in time evolution simulations. Although a seed with a single gravitational normal mode can already trigger secular resonances, their frequency is always the same as the normal mode frequency we start with. The smoking gun of a frequency cascade is given by the appearance of secular resonances at third order that have frequencies different from the seed. This requires starting with a seed that is the superposition (collision) of at least two normal modes. As an example, take the seed:
 \begin{eqnarray}
&& \hspace{-1cm} \{ \ell, m,p, \bar{\omega} \}_{\rm \bf s} =\{4,4,0,5/L\}, \quad \hbox{amplitude}\:\:\mathcal{A}^{(1)}_{({\rm \bf s}) 4} \varepsilon, \nonumber\\ 
&& \hspace{-1cm} \{ \ell, m,p, \bar{\omega} \}_{\rm \bf s} =\{6,6,0,7/L\}, \quad  \hbox{amplitude}\:\:\mathcal{A}^{(1)}_{({\rm \bf s}) 6}\varepsilon, 
\label{initialdata}
\end{eqnarray}
where $\mathcal{A}^{(1)}_{({\rm \bf s}) 4}$ and $\mathcal{A}^{(1)}_{({\rm \bf s}) 6}$ are $\mathcal{O}(1)$ quantities.

At second order, 15 scalar and 10 vector harmonics are excited and the solution can be made asymptotically global AdS and regular without resonances. Third order excites a total of 30 scalar and 22 vector harmonics. There are two resonances, $\{ \ell, m,p, \omega \}_{\rm \bf s}= \{4,4,0,5/L \}$ and  $\{ \ell, m,p, \omega \}_{\rm \bf s}=\{6,6,0,7/L \}$, which can be removed using a 
Poincar\'e-Lindstedt frequency correction \eqref{O2:wexpansion} with $\omega_{4,4,0}^{(2)}$ and  $\omega_{6,6,0}^{(2)}$ given by the values in Table \ref{Table:geonsS1}.  But we also have two secular irremovable resonances,  
\begin{equation}
\{ \ell, m,p, \omega \}_{\rm \bf s} ={\bigl \{} \{2,2,0,3/L \}, \{8,8,0,9/L \} {\bigr \}} ,
\label{O3:irremovableR}
\end{equation}
whose quantum numbers do not coincide with \eqref{initialdata}. The first mode in  \eqref{O3:irremovableR} has $\omega=\frac{3}{L}\equiv\bar{\omega}_{\ell_{\rm \bf s}=2, p=0}$. From \eqref{spectrumS} this is a normal mode frequency {\it smaller} than the two frequencies of the seed \eqref{initialdata}. On the other hand, the second mode in  \eqref{O3:irremovableR} has $\omega=\frac{9}{L}\equiv\bar{\omega}_{\ell_{\rm \bf s}=8, p=0}$ which is a normal mode frequency {\it larger} than any of the two frequencies contained in the seed \eqref{initialdata}. This shows that the weakly perturbative turbulent mechanism predicts the generation of irremovable resonances that have {\it both} larger and smaller frequency than those present in the seed. This signals that the time evolution of such a seed should proceed with a {\it direct and inverse} cascades of frequencies.

In an attempt to understand which of the cascades is likely to dominate faster the time-evolution, we have compared the coefficient of the direct and inverse cascades. In order to do this in a gauge invariant way, we computed the boundary holographic stress energy tensor \cite{Balasubramanian:1999re,deHaro:2000vlm} and its concomitant energy density. We then compared the ratio between the two secular terms. If we assume that each of the modes in the seed carries {\it equal energy}, the direct cascade is a factor of 10 larger than the inverse cascade, perhaps signalling that black hole formation is likely to occur at late times.

%%%%%%%%%%%%%%%%%%%%%%%%%%%%%%%%%%%%%%%%%%%%%%%%%%
%\section{4.~Direct and inverse turbulent cascades:}
%%%%%%%%%%%%%%%%%%%%%%%%%%%%%%%%%%%%%%%%%%%%%%%%%%
{\bf 5.~Gravitational sector is more populated by secular resonances ---}
In this  section we highlight that gravitational perturbations about AdS have a much richer structure than in the scalar field sector. On one hand, unlike the scalar field case, this is because seeds with a single mode can already develop irremovable resonances (Section 3). In addition, there is an enhancement on the number of irremovable resonances if we start with two or more normal modes of AdS.

To illustrate this, take for seed the combination: 
 \begin{eqnarray}
&& \hspace{-1cm}  \{ \ell, m,p, \bar{\omega} \}_{\rm \bf s}=\{4,4,0,5/L\}, \quad  \hbox{amplitude}\:\:\mathcal{A}^{(1)}_{({\rm \bf s}) 4} \varepsilon\,; \nonumber\\ 
&& \hspace{-1cm}   \{ \ell, m,p, \bar{\omega} \}_{\rm \bf v}=\{7,6,0,9/L\}, \quad \hbox{amplitude}\:\:\mathcal{A}^{(1)}_{({\rm \bf v}) 7}\varepsilon\,. 
\label{initialdata2}
\end{eqnarray}
The first is a scalar mode that in isolation does not develop irremovable resonances, while the second is a vector mode that does so: see Tables \ref{Table:geonsS1} and \ref{Table:geonsV}.

At second order, there are 16 scalar and 13 vector harmonics excited and the solution can be made asymptotically global AdS and regular without introducing resonances. At third order, a total of 34 scalar and 31 vector harmonics are excited. There are two resonances, namely $\{ \ell, m,p, \omega \}_{\rm \bf s}= \{4,4,0,5/L\}$ and  $\{ \ell, m,p, \omega \}_{\rm \bf v}=\{7,6,0,9/L\}$ which we remove using a Poincar\'e-Lindstedt frequency correction \eqref{O2:wexpansion}, 
{\small
\begin{eqnarray}
&& \hspace{-0.4cm}  \omega_{4,4,0}^{(2)} L = -\frac{7010569125}{77792}-\frac{1860284480041845}{7607296}\frac{(\mathcal{A}^{(1)}_{({\rm \bf v}) 7})^2}{(\mathcal{A}^{(1)}_{({\rm \bf s}) 4})^2}, \label{O3:wCorrections}\nonumber\\
&& \hspace{-0.4
cm}  \omega_{7,6,0}^{(2)} L= -\frac{8548214990390361}{19124224}-\frac{21681637296525}{271960832} \frac{(\mathcal{A}^{(1)}_{({\rm \bf s}) 4})^2}{(\mathcal{A}^{(1)}_{({\rm \bf v}) 7})^2}.  \nonumber
\end{eqnarray} 
}
Here, the first contribution to each frequency matches the values in Tables \ref{Table:geonsS1} and \ref{Table:geonsV}. It removes a resonance of the type $2\bar{\omega}_4-\bar{\omega}_4=5/L$ and  $2\bar{\omega}_7-\bar{\omega}_7=9/L$, respectively. The second contribution is due to the interaction of the initial modes in \eqref{initialdata2} and is proportional to their relative initial amplitudes. It removes a resonance of the type $\bar{\omega}_4+\bar{\omega}_7-\bar{\omega}_7=5/L$ and $\bar{\omega}_7+\bar{\omega}_4-\bar{\omega}_4=9/L$, respectively.

In addition we have 7 irremovable secular resonances with quantum numbers found in the normal mode spectra \eqref{spectrumS} and  \eqref{spectrumV}, but that do not coincide with the data \eqref{initialdata2}, 
\begin{eqnarray}
&&\hspace{-1cm} \{ \ell, m,p, \omega \}_{\rm \bf s} = {\bigl\{ }  \{6,6,1,9/L \}, \{8,6,0,9/L \}{\bigr \} };  \nonumber\\
&& \hspace{-1cm}\{ \ell, m,p, \omega \}_{\rm \bf s} =  {\bigl\{ } \{8,8,2,13/L \}, \{10,8,1,13/L \}, \nonumber\\ 
&&\hspace{1.55cm}\{12,8,0,13/L \}  {\bigr \} };  \nonumber\\
&& \hspace{-1cm}\{ \ell, m,p, \omega \}_{\rm \bf v} =  {\bigl\{ }  \{9,8,1,13/L \}, \{11,8,0,13/L \}  {\bigr \} }. \label{O3:irremovableR2}
\end{eqnarray}
Here, the first line resonances are of the type $2\bar{\omega}_7-\bar{\omega}_7=9/L$ and  coincide with those in the last column of Table \ref{Table:geonsV} when we start with the single normal mode  $\{ \ell, m,p, \bar{\omega} \}_{\rm \bf v} =\{7,6,0,9/L\}$. However, the second and third line resonances are of the type $2\bar{\omega}_7-\bar{\omega}_4=13/L$ and thus a consequence of the two-mode collision \eqref{initialdata2}.

If we collide two geons, like in \eqref{initialdata}, we get only two irremovable secular resonances. But when one of the starting modes $-$ like \eqref{initialdata2} $-$ or both do not have a geon extension, more than two secular resonances are generated. This is unique to the gravitational sector. 

%%%%%%%%%%%%%%%%%%%%%%%%%%%%%%%%%%%%%%%%%%%%%%%%%%
%\section{6. Discussion:}
%%%%%%%%%%%%%%%%%%%%%%%%%%%%%%%%%%%%%%%%%%%%%%%%%%
{\bf 6.~Final discussions ---} The educated and judiciously chosen examples that we have analysed in Tables  \ref{Table:geonsS1} and \ref{Table:geonsV} (and others not shown) give strong evidence to the following conjecture.
{\it The only gravitational normal modes of AdS$_4$ that can be back-reacted up to third order to yield a time-periodic horizonless solution of the Einstein equation are:}

$\bullet \,$ Scalar modes with $\ell_{\rm s}=|m_{\rm s}|\geq 2$ and $p_{\rm s}=0$, or 

$\bullet \,$ Scalar modes with $\ell_{\rm s}= 2$, $m_{\rm s}=0,1$ and $p_{\rm s}=0$, or 

$\bullet \,$ Vector modes with $\ell_{\rm v}= 2$, $m_{\rm v}=0$ and $p_{\rm v}=0$.

We have further explicitly check that the $\{\ell_{\rm s},m_{\rm s},p_{\rm s}\}=\{2,2,0\}$ case can be back-reacted up to $\mathcal{O}(5)$ and actually up to any order, since it has a full nonlinear extension \cite{Dias:2011ss,Horowitz:2014hja}. This fact, and the observation that the structure of the problem is the same for all $\ell_{\rm s}=|m_{\rm s}|\geq 2$ and $p_{\rm s}=0$ modes, allows to conjecture that the first case in our list can be nonlinearly extended to yield a geon. 

Having only a `few' normal modes with a solitonic extension is unique to the gravitational sector. Indeed, for a real (complex) scalar field \emph{any} normal mode can be back-reacted to yield an oscillon (boson star). 

Ultimately, this property is due to gravity having two fundamentally distinct sectors (scalar and vector) of normal modes whose frequency spectra \eqref{spectrumS} and  \eqref{spectrumV} depends on two quantum numbers (not one): $\ell$ and $p$. (A spherically symmetric scalar field has a single normal mode sector and its spectrum depends only on $p$). Consequently different combinations of $\{\ell,p\}$ can yield the same frequency and make the system prone to develop more secular resonances than in the spherically symmetric scalar field case, as confirmed in Tables \ref{Table:geonsS1} and \ref{Table:geonsV}. 

As a consequence of these properties, a collision of two (or more) gravitational normal modes that do not have a geon extension generates more than just a pair of irremovable resonances. This is illustrated by \eqref{initialdata2} and \eqref{O3:irremovableR2}. Had we replaced the first mode in  \eqref{initialdata2} by a `non-geonic' normal mode and we would generate even more irremovable resonances. Furthermore, in the spherically symmetry scalar field collapse, when we consider $n\geq2$ distinct initial seeds, we get $n^2(n-1)/2$ irremovable resonances at third order in perturbation theory. For the gravitational case we expect a much faster growth of the number of irremovable resonances with $n$, although its precise account is likely to depend on the vales of $(\ell,m)$ of each of the $n$ seeds. We take this as strong evidence suggesting that the time evolution of the gravitational nonlinear instability of AdS should differ considerably from its spherical symmetric scalar field counterpart. In a sense it should be more dramatic and possibly even faster (although our perturbation theory analysis still breaks down at third order and thus for timescales of order $t \geq \varepsilon^{-2}$).

Our analysis also has striking implications for the moduli space of rotating black hole solutions in AdS. In particular, the zero horizon radius limit of a (superradiant) `black resonator' \cite{Dias:2015rxy} is a geon only for the cases listed above. For any other choice of quantum numbers the zero horizon radius limit of black resonators is likely to be singular \footnote{Hairy black holes with singular zero horizon radius limit were found in \cite{Dias:2011tj}.}.

%\smallskip
%%%%%%%%%%%%%%%%%%%%%%%%%%%%%%%%%%%%%%%%%%%%%%%%%%%%%%%%%%%%%%%%%%%%%%%%%%
\vskip .5cm
\centerline{\bf Acknowledgements}
\vskip .2cm
We wish to warmly  thank Gary Horowitz for useful comments on an earlier version of this manuscript. O.D. acknowledges financial support from the STFC Ernest Rutherford grants ST/K005391/1 and ST/M004147/1. 
%O.D. thanks the Instituto de Fisica Teorica (IFT UAM-CSIC) in Madrid for its support via the Centro de Excelencia Severo Ochoa Program under Grant SEV-2012-0249, and the Workshop ?Iberian Strings 2016(iStrings)? where part of this work was developed.
\newpage

\onecolumngrid  \vspace{1cm} 
\begin{center}  
{\Large\bf Supplementary Material} 
\end{center} 
\appendix 
%\tableofcontents  
%%%%%%%%%%%%%%%%%%%%%%%%%%%%%%%%%%%%%%%%%%%%%%%%%%
\section{Statement of the perturbation theory problem}
%%%%%%%%%%%%%%%%%%%%%%%%%%%%%%%%%%%%%%%%%%%%%%%%%%
We consider Einstein-AdS theory in four dimensions with action and equation of motion, 
\begin{equation}
S=\int d^4x\;\sqrt{-g}\left(R+\frac{6}{L^2}\right), \qquad R_{\mu\nu}+\frac{3}{L^2}g_{\mu\nu}=0\,,
\end{equation}
where $L$ is the AdS length scale. Within perturbation theory, we perturb this equation of motion about the AdS background $\bar{g}$  as  $g = \bar{g}+\sum_k h^{(k)}\varepsilon^k$. Here, $\varepsilon$ is  a perturbation parameter whose physical meaning will be understood later, and $\bar{g}$ is the metric of global AdS,
\begin{equation}
\bar{g} = -\left(1+\frac{r^2}{L^2}\right)dt^2+\frac{dr^2}{1+\frac{r^2}{L^2}}+r^2(d\theta^2+\sin^2\theta d\phi^2).
\end{equation}
At each order  in perturbation theory, the Einstein equations yield \eqref{eq:perturb}, $\Delta_L h_{ab}^{(k)} = T^{(k)}_{ab}$, where $\Delta_L$ is a second order operator constructed uniquely from $\bar{g}$,
\begin{equation}
2\Delta_L h_{ab}^{(k)} \equiv -\bar \nabla^2 h_{ab}^{(k)}-2 \bar{R}_{a\phantom{c}b\phantom{d}}^{\phantom{a}c\phantom{b}d}h_{cd}^{(k)}-\bar{\nabla}_{a}\bar{\nabla}_b h^{(k)}+2\bar{\nabla}_{(a} \bar{\nabla}^c h^{(k)}_{b)c}.
\end{equation}
with $h^{(k)}\equiv \bar{g}^{ab}h_{ab}^{(k)}$, and $\bar{R}_{abcd}$ being the AdS Riemann tensor. 
$T^{(k)}$ is a function of $\{h^{(j\leq i-1)}\}$ and their derivatives. At second order $T^{(2)}$ reduces to the familiar Landau-Lifshitz pseudotensor \cite{Landau:1951}. As a consequence of the Bianchi identities, $\bar \nabla^a T^{(k)}_{ab} =0$ for each order $k$.

We are interested on finding regular finite energy and angular momentum solutions of (\ref{eq:perturb}) for $k\geq 1$. The $SO(3)$ symmetry of AdS can be used to show that any regular two-tensor, say $T$, can be written as a sum of two fundamental building blocks \cite{Kodama:2003jz}
\begin{equation}
\mathcal{T} = \sum_{\ell_s,m_s} \mathcal{T}^{(s)}_{\ell_s,m_s}+\sum_{\ell_v,m_v} \mathcal{T}^{(v)}_{\ell_v,m_v}+\cos \phi\leftrightarrow\sin \phi,
\label{eq:linearsol1}
\end{equation}
where $\mathcal{T}^{(s)}$ ($\mathcal{T}^{(v)}$) represent scalar (vector) type modes. They are symmetric two-tensors built from scalar (vector) harmonics on the two-sphere $S^2$. Here $(\ell,m)$ are the standard labels for spherical harmonics, and vector harmonics on $S^2$ are of the form ${}^*\nabla Y_{\ell,m}$. Because we are using a real representation for the spherical harmonics, we need to account for both  the $\sin \phi$ and $\cos \phi$ terms and this justifies the last contribution in (\ref{eq:linearsol1}). Each $A^{(s)}_{\ell_s,m_s}$  ($A^{(v)}_{\ell_v,m_v}$) is parametrized by seven (three) arbitrary functions of $t$ and $r$.

Applying the expansion (\ref{eq:linearsol1}) to both $h^{(k)}$ and $T^{(k)}$ in (\ref{eq:perturb}), it follows that any solution of (\ref{eq:perturb}) is described by two decoupled PDEs of the form \eqref{eq:master}, $\Box_2 \Phi^{(k)}_{\ell,m}(t,r)+V^{(k)}_{\ell}(r) \Phi^{(k)}_{\ell,m}(t,r)=\widetilde{\mathcal{T}}^{(k)}_{\ell,m}(t,r)$, where $V^{(k)}_\ell(r)=- \ell  (\ell +1)/[r^2 (1+r^2/L^2)]$ is a potential that only depends on $\ell$ and $\Box_2$ is the d'Alambertian of the orbit space $ds^2 = -(1+r^2/L^2) dt^2+L^2 dr^2/(L^2+r^2)$ \cite{Kodama:2003jz}. One of the equations governs scalar-type modes and the other vector-type modes.
$\Phi^{(k)}_{\ell,m}(t,r)$ is a Kodama-Ishibashi (gauge invariant) variable from which  $h^{(k)}_{\ell,m}$ $-$ that solves (\ref{eq:perturb}) $-$ can be uniquely recovered (in a particular gauge)  through a linear differential map \cite{Kodama:2003jz}. Still in \eqref{eq:master}, $\widetilde{\mathcal{T}}^{(k)}_{\ell,m}(t,r)$ is a scalar source term, that can be expressed as a function of the components of $\mathcal{T}^{(k)}_{\ell,m}$ and its derivatives.

At each order in perturbation theory, we must impose regularity of  $h^{(k)}_{\ell,m}$, seen as a two-tensor on the fixed AdS background. This translates uniquely to a regularity condition on the associated master solution of (\ref{eq:master}). Namely, we must require $\Phi^{(k)}_{\ell,m}\sim \mathcal{O}(r^{\ell})$, as $r\to0$ for $h^{(k)}_{\ell,m}$ to be regular at the origin. At asymptotic infinity, We demand that the metric $g = \bar{g}+\sum_k h^{(k)}\varepsilon^k$ is asymptotically globally AdS. The source $\widetilde{\mathcal{T}}^{(k)}_{\ell,m}$ has an asymptotic decay such that any solution of (\ref{eq:master}) falls-off as
\begin{equation}
\Phi^{(k)}_{\ell,m}\sim A_{\ell,m}(t)+ \frac{B_{\ell,m}(t)}{r}+\mathcal{O}(r^{-2}),
\label{eq:asymp}
\end{equation}
for arbitrary functions  $A_{\ell,m}$ and $B_{\ell,m}$ of $t$. The asymptotically globally AdS boundary condition is imposed on $h^{(k)}_{\ell,m}$  \emph{not} on $\Phi^{(k)}_{\ell,m}$ directly. Thus, we first reconstruct $h^{(k)}_{\ell,m}$  and, in some cases, add a gauge transformation before imposing the boundary condition on $\Phi^{(k)}_{\ell,m}$. For scalar modes, the requirement that our geometry approaches asymptotically the Einstein static universe demands the boundary condition  $B_{\ell,m}(t)=0$. On the other hand the asymptotic boundary condition for vector modes is  $A_{\ell,m}(t)=0$.

A few fundamental observations are:  i)  to determine the source term in (\ref{eq:master}) we always need first to calculate all the previous order $h^{(k)}_{\ell,m}$;
ii) in our case the time dependence of the source term is an overall multiplicative factor of the form $\cos(\omega t)$, so (\ref{eq:master}) is solved assuming the separation {\it ansatz} $\Phi^{(k)}_{\ell,m}(t,r)=\cos(\omega  t) R(r)$. There is however an exception to ii), namely when (and only in this case) the source frequency $\bar{\omega}_{\ell,m}$ coincides with one of the AdS gravitational normal mode frequencies $\bar{\omega}$ in \eqref{spectrumS} or  \eqref{spectrumV}. If this is the case we say that the modes are \emph{resonant} and the general solution of (\ref{eq:master}) now includes a particular secular solution that grows with time:
\begin{equation}
\Phi^{(k)}_{\ell,m}(t,r)=\cos(\omega t)R_{\omega \ell,m}(r)+t\,\sin(\omega t)L_{\omega ,\ell,m}(r).
\label{eq:resonantBagain}
\end{equation}

%%%%%%%%%%%%%%%%%%%%%%%%%%%%%%%%%%%%%%%%%%%%%%%

\bibliography{refs}{}
\end{document}